\documentclass[doublecol]{epl2} 
\def\be{\begin{equation}}
\def\ee{\end{equation}}
\usepackage{psfrag}
\usepackage{graphicx}
\usepackage{amssymb}
\title{Fast decay of the velocity autocorrelation function in
dense shear flow of inelastic hard spheres}
\author{Ashish V. Orpe,$^{1,}$\footnote{Present Address: Chemical Engineering Division, National Chemical Laboratory, Pune 411008 India} V. Kumaran,$^2$ K. Anki Reddy,$^2$ and Arshad Kudrolli$^1$}
\institute{$^1$Department of Physics, Clark University, Worcester, MA 01610\\
$^2$Department of Chemical Engineering, Indian Institute of Science, Bangalore 560012 India} 

\pacs{47.57.Gc}{Granular flow}
\pacs{45.70.Mg}{Granular flow: mixing, segregation and stratification}

\abstract{
We find in complementary experiments and event driven simulations of sheared inelastic hard spheres that the velocity autocorrelation function $\psi(t)$ decays much faster than $t^{-3/2}$ obtained for a fluid of elastic spheres at equilibrium. Particle displacements are measured in experiments inside a gravity driven flow sheared by a rough wall. The average packing fraction obtained in the experiments is 0.59, and the packing fraction in the simulations is varied between 0.5 and 0.59. The motion is observed to be diffusive over long times except in experiments where there is layering of particles parallel to boundaries, and diffusion is inhibited between layers. Regardless, a rapid decay of $\psi(t)$ is observed, indicating that this is a feature of the sheared dissipative fluid, and is independent of the details of the relative particle arrangements. An important implication of our study is that the non-analytic contribution to the shear stress may not be present in a sheared
inelastic fluid, leading to a wider range of applicability of kinetic
theory approaches to dense granular matter.
}
\begin{document}

\maketitle

% References should be done using the \cite, \ref, and \label commands

A significant feature of the dynamics of dense fluids is the effect of correlations on transport coefficients~\cite{ernst,kg}. For a gas of elastic particles in the dilute limit, the transport coefficients are calculated using the molecular chaos approximation~\cite{cc}, that the two-particle velocity distribution is the product of the single particle velocity distribution functions. It is known that this procedure cannot be extended to higher densities due to the effect of correlations. For a $D$ dimensional elastic fluid, the  velocity autocorrelation function $\psi(t)$ decays as $t^{-D/2}$,  and this power law decay is  referred to as the ``long time tail" in the velocity autocorrelation function~\cite{aw}. This was recognized as a consequence of the diffusive transport of momentum proportional to the square of the wave vector in a system in which momentum and energy are conserved. This leads to a divergence of the viscosity in two dimensions. In three dimensions, there is a non-analytic contribution to the shear stress proportional to $\dot{\gamma}^{3/2}$ in limit of zero shear rate~\cite{ernst}; this correction is larger than the Burnett terms (proportional to $\dot{\gamma}^2$ in the constitutive relation for the shear stress obtained by the Chapman-Enskog procedure. These correlations have been anticipated to impact extention of kinetic theory to dense granular flows since its first application to dilute granular gases, and have been also seen in uniform granular flows~\cite{orpe07}. Because dense flows are more prevalent, such observations have led to the belief that kinetic theory approaches are severely limited. 

However, more recently, it has been speculated that the decay of $\psi(t)$ in sheared granular flows could be faster~\cite{oh,vk}. For a fluid under shear, a steady state can be obtained only when there is an energy dissipation mechanism, since there is a continuous increase in the fluctuating (thermal) energy of the fluid particles, and the rate of increase is equal to the product of the stress and the strain rate. The two most widely studied mechanisms are thermostats (where there is a drag force on the particles) and inelastic collisions. In a sheared elastic fluid in the absence of thermostats, there is a continuous increase in the thermal energy (temperature) of the particles. Due to this, it is not possible to calculate the decay of velocity autocorrelation functions at steady state; the decay of the autocorrelation function will depend both on the initial temperature and the rate of heating, and it will not be invariant with respect to translation in time. The steady state is completely defined only if we specify both the energy production mechanism (mean shear) and the energy dissipation mechanism (inelastic collisions or thermostat).

The velocity autocorrelation function in a thermostated system with inelastic collisions and random noise has been examined by Puglisi et al~\cite{puglisi}, and they have found that the decay is not exponential, but rather has a behaviour similar to a stretched exponential. This is in contrast to the algebraic decay in an elastic fluid at equilibrium in the absence of shear~\cite{aw}. The reason for this difference could be the following. In an elastic fluid, the long time tails in the velocity correlation functions is due to the diffusive transport of the transverse momentum in the fluid~\cite{ernst}. In a thermostated system, momentum is not a conserved variable locally, since there are drag and fluctuating forces acting on the individual particles. This could result in a faster decay in the velocity autocorrelation functions.

Here, we analyse a class of sheared dissipative fluids with inelastic collisions, where momentum is conserved in collisions, and energy is dissipated. Due to this, energy is now a non-conserved variable, and we examine the effect of mean shear and energy dissipation on the velocity autocorrelation function. It has been suggested~\cite{oh,vk} that the velocity autocorrelation function in a sheared inelastic fluid should decay much faster than the $t^{-D/2}$ power law decay in an elastic fluid at equilibrium.

Kumaran~\cite{vk} incorporated correlations using the ring-kinetic equation for a dense gas to obtain a faster decay proportional to $t^{-9/2}$ in three dimensions. A later more detailed calculation~\cite{vk2}, which accounted for the rotation of the wave vector with the mean shear for calculating the wave-number cut-offs for perturbations modulated in the flow direction, indicated that the decay rate of the velocity autocorrelation function should be anisotropic, and should scale as $t^{- 15/4}$ in the flow plane, and as $t^{-7/2}$ in the vorticity direction. These results are in contrast  with the power law decay $t^{-5/2}$ obtained by Otsuki and Hayakawa~\cite{oh} in three dimensions using mode coupling theory. Either of these corrections would render the viscosity finite in two dimensions, and there is no non-analytic contribution to the viscosity proportional to $\dot{\gamma}^{3/2}$ in three dimensions, suggesting that calculations based on kinetic theory would have a {\em greater} range of validity, at least upto Burnett order in both two and three dimensions, for granular flows in comparison to elastic fluids.

We investigate the velocity autocorrelation function in a sheared granular flow in complementary experiments and simulations. In experiments, we produce a sheared granular flow with a rough boundary, and measure particle motion in three dimensions away from side walls using an interstitial liquid index matching technique. Further, to decouple the effect of the boundary on the correlations, to examine effect of volume fraction, and to examine a simpler granular system undergoing linear shear, we perform event driven (ED) simulations of inelastic spheres with Lees-Edwards boundary conditions. Therefore, in the experimental system the flow occurs past boundaries which are held fixed and the granular volume fraction can vary within the flow. In contrast, the simulations are performed with fixed volume fraction conditions. Independent of these details, the experiments and simulations show rapid decay of autocorrelations in a dense packing regime with volume fraction as high as 0.59. Further, the particles in both experiments and simulations do not show ordering in the plane perpendicular to the direction of shear, as evidenced by the in-plane hexagonal order parameter. Layering occurs near the boundaries in the experiments, while there is no order even in the gradient direction in the simulations because we use periodic boundary conditions. Despite these differences,  $\psi(t)$ shows a decay much faster than $t^{-3/2}$ in both cases. At the onset, we note that because of considerable challenges in measuring and tracking experimental particles in the bulk, the exact exponent is beyond our capability as it requires tracking particles longer than they are in our system. 

\begin{figure}
\includegraphics[width=0.95\linewidth]{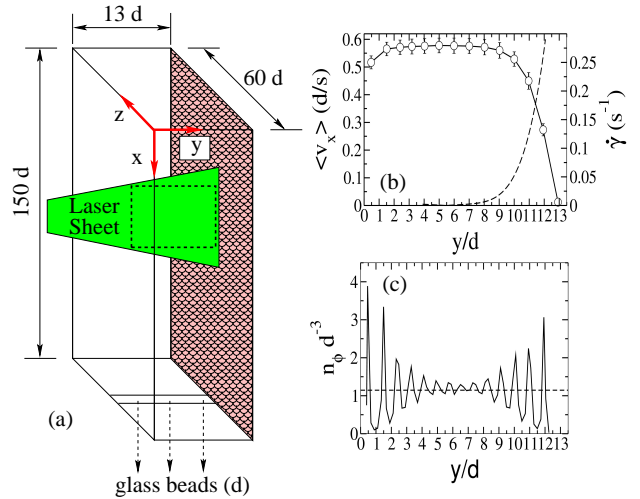}
\caption{(a) Schematic diagram of the granular flow cell used in the  experiments. A layer of glass beads is glued to the right side wall in order to roughen it. The observation area is indicated by the dashed box. (b) The mean velocity and the strain rate $\dot{\gamma}$ across the shear plane. (c) The number of particles per unit volume across the shear plane shows layering near the side walls.}
\label{schem}
\end{figure}

The experimental system consists of a vertical glass pipe with a rectangular cross section filled with glass beads with diameter $d = 1.0 \pm 0.1$\,mm, and density $\rho_g = 2.5 \times 10^{3}$ kg m$^{-3}$ driven by gravity. A layer of glass beads is glued to one of the sides to roughen it as shown in Fig.~\ref{schem}(a). The interstitial space between the grains is filled with a liquid with the same refractive index as the glass beads~\cite{tsai03,siavoshi06}. The liquid has density $\rho_l =  1.0 \times 10^{3}$ kg m$^{-3}$, and viscosity $\nu = 2.2 \times 10^{-2}$ kg m$^{-1}s$. A dye added to the liquid is illuminated by a light sheet and imaged from an orthogonal direction. Thus the particles appear dark against a bright background, and a centroid algorithm  is used to find the position of the particles to within a twentieth of a particle diameter. By switching the light and image planes, we can measure the particle motion in the shear and in the vorticity planes.

\begin{figure}
\includegraphics[width=0.95\linewidth]{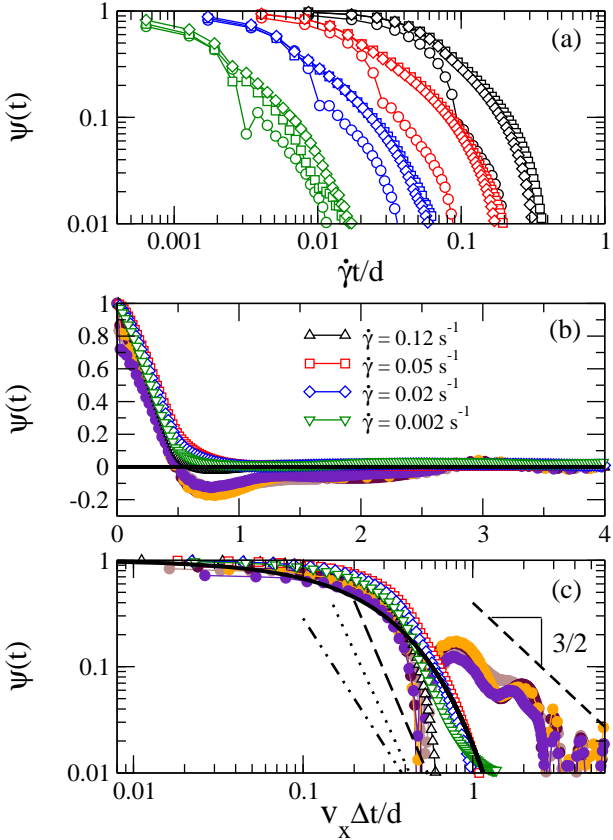}
\caption{(a)  Velocity auto-correlations function  $\psi_{ii}$ versus time scaled with strain rate obtained from the experiments. ($i = x - \square$, $y - \circ$, $z - \diamond$). $\psi_{ii}$ is observed to become more isotropic (right to left) as one moves away from the shearing wall where particles form layers. The four layers are centered around $1.57d$, $2.54d$, $3.49d$ and $4.48d$ from the rough wall. $\psi(t)$  measured in the experiment plotted in linear scale (b) and logscale (c). $\psi(t)$ decay faster in sheared flow (open symbol) compared with unsheared flows at various flow rates in Ref.~\cite{orpe07} (filled symbol). Lines with slope $-5/2 $ (dash dot), $-7/2$ (dot) and $-15/4$ (dash) are also added to guide the eye. An exponential fit (solid line) is also added for comparision.      
}
\label{vautoexpt}
\end{figure}

Figure~\ref{schem}(b) shows the measured mean flow velocity $<v_x>$ as a function of horizontal distance $y$ in the shear plane. To minimize the effect of the front and back walls, the data is obtained for $20d < z <40d$. The rough boundary is observed to slow the flow compared to the smooth boundary. At these measured velocities, the Reynolds number $Re = v_x d/\nu$ is about $10^{-2}$, and the ratio of the viscous drag of the grain to the gravitational force is estimated to be less than $10^{-2}$. The strain rate $\dot{\gamma}$ given by the gradient of the mean velocity is shown in Fig.~\ref{schem}(b) as well. The maximum strain rate $0.25 {\rm s}^{-1}$, is comparable in magnitude to a previous granular shear study~\cite{siavoshi06} where the drag friction experienced by the boundary was measurably unaffected by the presence of the same interstitial liquid. Thus we conclude that the lubrication forces between grains do not affect the grain motion in the region where we discuss the correlation properties. The side walls also impose layering as can be observed from the oscillations in the plot of number of particles in a volume $d^3$ plotted across the horizontal crossection in Fig.~\ref{schem}(c). The oscillations decrease with distance from the boundaries, and can be used to measure any influence of grain order on correlations. By taking the average of $n_\phi d^{-3}$ and multiplying by the volume of sphere, we obtain that the average volume fraction of the grains $\phi$ in the observation window is $0.59$. 

ED simulations are performed in a three dimensional box with 500 smooth inelastic spheres of diameter $d$ subjected to a rate of deformation field ${\bf G} = \dot{\gamma} {\bf e}_x {\bf e}_y$. The co-ordinate system is similar to that in the experiments, and the flow is in the $x$ direction, the velocity gradient in the $y$ direction and the $z$ co-ordinate is perpendicular to the flow plane. The box size is adjusted to obtain the desired volume fraction. Simulations in three dimensions at large volume fractions (upto 0.6) and with coefficient of restitution as low as 0.7 have been carried out using the ED method by several authors~\cite{mitarai}. In the ED procedure, particle trajectories are extrapolated forward in time and the earliest possible collision is identified. All particle positions are advanced by this time interval, and the  post-collisional velocity of the colliding particles is updated along the line joining centers of the particles with a factor $- e_n$ times the pre-collisional relative velocity, where $e_n$ is the  coefficient of restitution. The simulations were initiated with the particles arranged in an FCC lattice, and with velocities chosen from a random distribution with variance $1$. The simulation was first allowed to proceed for $2 \times 10^4$ collisions per particle in order to reach a steady state, and the averaging was carried out over another $2 \times 10^5$ collisions per particle. Shear flow is imposed using Lees-Edwards periodic boundary conditions. More details on the simulation methods can be found in Ref.~\cite{kumaran08}. For an inelastic fluid under shear, the strain rate $\dot{\gamma}$ and the `granular temperature' $T$ (mean square of the fluctuating velocities of the particles) are related through the energy balance condition, that the rate of shear production and inelastic dissipation of energy of the particles are equal at steady state. We carry out simulations with smooth spheres with  $e_n = 0.9$, and with the volume fraction up to $(\phi = 0.59)$. For these parameters, we have taken care to ensure that there is no inelastic collapse~\cite{alamhrenya}, and there are no particle overlaps due to repeated collisions between particles. In addition, we have verified from the profiles of the volume fraction and mean velocity that there is no clustering instability~\cite{hopkinslouge} for the system sizes and volume fractions considered here. In order to contrast our results to the equilibrium case, simulations are also performed with elastic particles in the absence of applied shear at the same volume fractions.   

\begin{figure}
\includegraphics[width=0.95\linewidth]{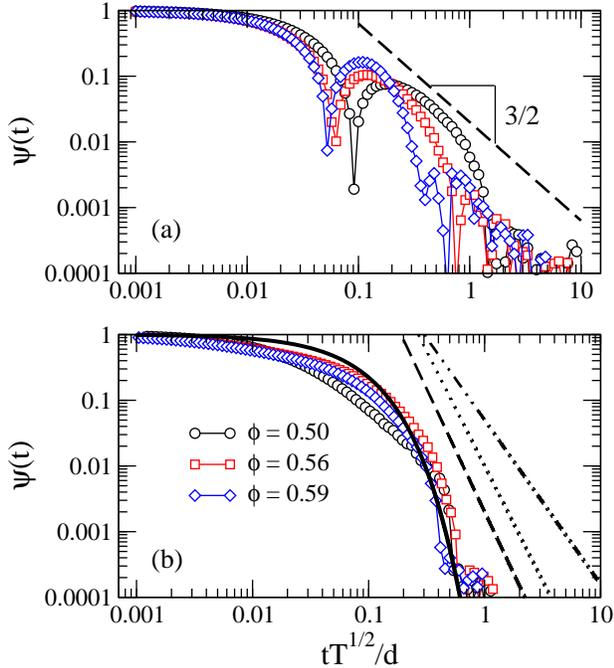}
\caption{(a) $\psi(t)$ obtained from simulations with elastic spheres at equilibrium. (symbols: $\phi= 0.50 - \circ$, $0.56 - \square$, $0.59 - \diamond$) $\psi(t)$ decays consistent with $t^{-3/2}$ for elastic case.     
(b) $\psi(t)$ obtained from simulations with sheared inelastic spheres. %(symbols: $\phi= 0.50 - \circ$, $0.56 - \square$, $0.59 - \diamond$). 
$\psi(t)$ for sheared inelastic spheres decays faster than $t^{-3/2}$. Lines with slope $-5/2 $ (dash dot), $-7/2$ (dot) and $-15/4$ (dash) are also added to guide the eye. An exponential fit (solid line) is also added for comparision.  
}
\label{vautosim}
\end{figure}

Figure~\ref{vautoexpt}(a) shows $\psi(t)$ defined as,
\be
\psi_{ij}(t) = \frac{\langle \Delta u_i(t) \Delta u_j(0) \rangle}{\langle 
\Delta u_i(0) \Delta u_j(0) \rangle}
\ee
where $\Delta {\bf u}$ is the fluctuating velocity of the 
particles, and $i$ and $j$ correspond to the directions $x$, $y$, or $z$, and the average is over all the particles tracked over all times in that direction. $\psi(t)$ corresponding to the experiments shown in Fig.~\ref{vautoexpt}(a), is measured by calculating $\Delta {\bf u}$ from the displacement of the particle over a time interval over which the mean flow travels a distance of $0.1d$ at that location. In the experiments, each particle is tracked as long as its center is within the layer ($1d$ wide) in the shear direction so has to have strain rate approximately constant. We terminate the trajectory even if the particle leaves the layer momentarily. Therefore if it returns back to the same layer at a later time then it is considered as a new trajectory. $\psi(t)$ clearly decays rapidly over a time scale less than $1/\dot{\gamma}$. It is also observed that $\psi_{yy}$ decreases more rapidly in the boundary regions where particles are layered.  $\psi(t)$ would collapse on to one curve if $T$ was uniform in the system. 
Because there is an energy current in the experiments, and $T$ is not exactly proportional to $\dot{\gamma}^2$, multiple curves are observed. 

Next, in order to contrast with correlations observed in unsheared granular flows, we replot $\psi(t)$ for the sheared case with the corresponding  $\psi(t)$ for uniform (unsheared) granular flow obtained at various flow rates published previously in Ref.~\cite{orpe07}. Here, we have averaged over the $x$ and $z$ directions to obtain smoother curves because $\psi(t)$ appears to be isotropic in the vorticity plane.  Scaled this way, $\psi(t)$ collapse on to two sets of curves depending on the presence or absence of shear\footnote{It may be more natural to scale with the square root of the granular temperature, but at such high densities, it is impossible to obtain instantaneous velocities of particles in the experiment.}. In case of the unsheared case, $\psi(t)$ crosses zero and becomes negative before decaying to zero for all the curves corresponding to different flow rates (see Fig.~\ref{vautoexpt}(b)). Whereas  $\psi(t)$ for the sheared case decays differently and faster to zero than $t^{-3/2}$. The same data is plotted in log-log scale in Fig.~\ref{vautoexpt}(c) to show case the fast decays further. It may be noted that the decay is closer to the exponential fit plotted than to the unsheared case. While lines with slope $-15/4$, $-7/2$~\cite{vk2} and $-5/2$~\cite{oh} are drawn to guide the eye, these decays are so fast that the range over which $\psi(t)$ can be measured does not allow us to draw any firm conclusions on the scaling of the decay. 

In Fig.~\ref{vautosim}, $\psi(t)$ obtained from simulations of inelastic sheared spheres with $e_n = 0.9$ is shown, along with simulations of elastic spheres at the same $\phi$. Decay proportional to $t^{-3/2}$ is clearly observed for low volume fractions $\phi = 0.2$, which is not shown in Fig.~\ref{vautosim}(a) for sake of clarity. For $\phi = 0.4$ and higher, it is well known that there is a reversal in the sign of $\psi(t)$ from positive to negative at a finite time, due to the reversal in the velocity of the particle trapped in the cage between its neighbors. The time for this reversal decreases as $\phi$ increases (as can be noted from the shift of the negative peak in Fig.~\ref{vautosim}(a)), due to a decrease in the dimensions of the cage formed by the neighboring particles. However, the absolute  value of $\psi(t)$ still shows a power law decay proportional to $t^{-3/2}$ in the long time limit. By contrast, the sheared simulation data does not show this reversal and shows a fast time decay for all $\phi$. Interestingly, it may be noted that $\psi(t)$ in the three directions are almost equal to each other except when layering is important.

To examine any link between orientational order and $\psi(t)$, we examine the two dimensional orientational order parameter $q_6 = <\exp{(i6\theta_p)}>$,
where $<.>$ is the average over all nearest neighbor bonds and $\theta_p$ is the angle between particle bonds. For disordered packing, $q_6 = 0$. Evaluating $q_6$ for the experimental data in the vorticity plane, we find that $q_6$ decreases from about 0.075 near the side walls to less than 0.03, a few grain diameters away from the side wall, indicating disordered packing. $q_6$ in the shear plane is observed to be $\sim 0.1$ but still significantly smaller than $q_6 = 1$ for hexagonal order. $q_6$ obtained from the simulations, both in the vorticity plane and the shear plane, is also observed to be less than $0.03$, indicating that there is no hexagonal ordering. 

\begin{figure}
\includegraphics[width=0.9\linewidth]{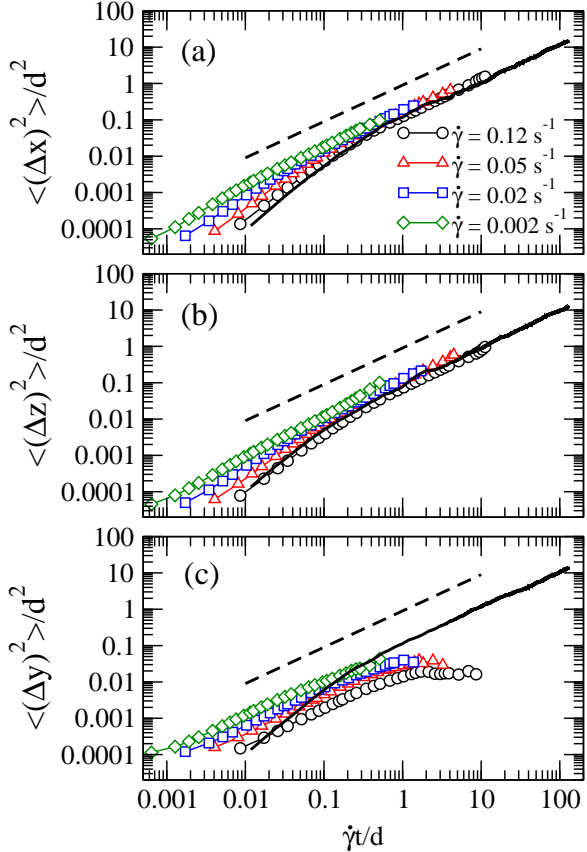}
\caption{The scaled mean square displacements (MSD) in (a) the flow direction, (b) the vorticity direction, and (c) the shear direction in the experiments (symbols) and the simulations (solid line) for various strain rates as a function of the scaled time $t \dot{\gamma}$. The contribution of the affine deformation of the flow has been subtracted from the displacements (see text). The dashed line corresponds to a slope of $1$. 
}
\label{msd}
\end{figure}

However, the state of order appears to be linked to particle diffusion in a shear flow of inelastic particles, as quantified by the mean square displacement (MSD) of the particles (see  Fig.~\ref{msd}.) To calculate MSD in a shear flow, one has to take into account the affine deformation of the flow (sometimes also refereed to as Taylor dispersion~\cite{utter04}.) Therefore, $\langle (\Delta x)^2 \rangle = \langle (x(t) - x^a(t))^2 \rangle$, $\langle (\Delta y)^2 \rangle = \langle (y(t) - y(0))^2 \rangle$, and $\langle (\Delta z)^2 \rangle = \langle (z(t) - z(0))^2 \rangle$, where  $x^a(t)$, corresponds to the affine deformation of the shear flow. After accounting for this contribution, we find that slope approaches 1 over long times in the vorticity and the flow directions, showing that the diffusion constants scale with strain rate. However in the gradient direction, MSD in the experiments appear to grow sub-diffusively and particles even appeared trapped in a layer over the observation window. This effect is most pronounced in the region next to the side wall, and the slope only appears to approach 1, as one examines regions away from the side walls. Because of the Lees-Edwards boundary conditions used in the simulations, there is no layering, and a clear crossover to diffusive behavior is observed even in the gradient direction for $\dot{\gamma} t/d > 1$.  

Therefore, it appears that the rapid decay of $\psi(t)$ is not related to the sub-diffusive nature of the MSD (caging effect), but appears to be present independent of the state of order and the diffusive/sub-diffusive nature of particle motion. These results clearly show that the velocity autocorrelation function decays much faster than the $t^{-3/2}$ decay for a fluid of elastic particles both in the presence and absence of ordering, indicating that it seems to be an inherent property of a dissipative sheared system. 

Finally, we argue that even though the experiments are done with particles immersed in a fluid, the very small ratio of the viscous and the gravitational forces means than the momentum is essentially transfered between particles during contact and very little is transferred to the interstitial fluid just as in dry granular matter colliding in air. Therefore, we anticipate that the analysis performed with ring-kinetic approach~\cite{vk} carry over because momentum is conserved and energy is dissipated except perhaps in the short time limit depending on the details of the interaction.

In summary, we have shown that the decay of the velocity autocorrelation function in a sheared granular flow is qualitatively different from that in an elastic fluid at equilibrium. Our results cannot verify the scaling laws~\cite{oh,vk}, because one would need to track the decay in the 
autocorrelation function over more than two to four orders of magnitude 
for one magnitude of increase in the time. However, the fast decay in the
velocity autocorrelation function indicates that the dynamics
in a sheared granular flow is different from that in an elastic
fluid at equilibrium. Specifically, {\em the
non-analytic contribution to the shear stress proportional to
$\dot{\gamma}^{3/2}$, which is present in an elastic fluid due to long time
tails in the autocorrelation functions,} may not be present in
a sheared inelastic fluid. This raises the prospect that kinetic
theory calculations may be applicable for a larger range of volume
fractions than equivalent theories for fluids of elastic particles.
Therefore, constitutive relations based on kinetic theory for 
granular flows may not be restricted to the dilute limit, as
previously thought, but may be applicable even at higher
volume fractions.

\begin{acknowledgments}
This work was supported by the National Science Foundation under Grant Nos.
 CTS-0334587, DMR-0605664 and the J.C. Bose Fellowship, DST, Government of India.
\end{acknowledgments}

% Create the reference section using BibTeX:
\bibliographystyle{apsrev}
%\bibliography{ak-bib}

\begin{thebibliography}{99}
\bibitem{ernst}M.H. Ernst, B. Cichocki, J.R. Dorfman, J. Sharma, 
\& H. van Beijeren, {J. Stat. Phys.}, {\bf 18}, 237 (1978).
\bibitem{kg}K. Kawasaki and J. Gunton, Phys. Rev. A {\bf 8}, 2048 (1973).
\bibitem{cc}S. Chapman and T.G. Cowling, {\em The Mathematical
Theory of Non-Uniform Gases}, Cambridge University Press, 1970.
\bibitem{aw}B. Alder and A.W. Wainwright,
{ Phys. Rev. A} {\bf 1}, 18 (1970).
\bibitem{orpe07}A.V. Orpe and A. Kudrolli, {Phys. Rev. Lett.} {\bf 98}, 238001 (2007).
\bibitem{oh}M. Otsuki and H. Hayakawa,
 cond-mat arXiv:0711.1421.
\bibitem{vk}V. Kumaran, {Phys. Rev. Lett.}, {\bf 96},
258002 (2006).
\bibitem{vk2} V. Kumaran, Phys. Rev. E, in press.
\bibitem{puglisi} A. Puglisi, A. Baldassarri and A. Vulpiani, 
J. Stat. Phys., P08016 (2007).
\bibitem{tsai03} J.-C. Tsai, G.A. Voth, and J.P. Gollub, Phys. Rev. Lett. {\bf 91}, 064301 (2003).
\bibitem{siavoshi06} S. Siavoshi, A.V. Orpe, and A. Kudrolli, Phys. Rev. E {\bf 73}, 010301 (2006).
\bibitem{mitarai}N. Mitarai and H. Nakanashi,
{Phys. Rev. E}, 031305 (2007);
C.S. Campbell, J. Fluid Mech. {\bf 348}, 85 (1997).
\bibitem{kumaran08} V. Kumaran, {\em submitted to} J. Fluid Mech.
\bibitem{alamhrenya} M. Alam and C.M. Hrenya, 
{Phys. Rev. E}, {\bf 63}, 061308 (2001).
\bibitem{hopkinslouge}M.A. Hopkins and M.-Y. Louge, 
{Phys. Fluids A}, {\bf 3}, 47 (1991).
\bibitem{utter04} B. Utter and R.P. Behringer, Phys. Rev. E {\bf 69}, 031308 (2004).
\end{thebibliography}

\end{document}